\documentstyle[12pt]{article}
\begin{document}
\def\beq{\begin{equation}}
\def\eeq{\end{equation}}
\def\bea{\begin{eqnarray}}
\def\eea{\end{eqnarray}}
\def\ve{\vert}
\def\vel{\left|}
\def\ver{\right|}
\def\nnb{\nonumber}
\def\ga{\left(}
\def\dr{\right)}
\def\aga{\left\{}
\def\adr{\right\}}
\def\rar{\rightarrow}
\def\nnb{\nonumber}
\def\la{\langle}
\def\ra{\rangle}
\def\lla{\left<}
\def\rra{\right>}
\def\ba{\begin{array}}
\def\ea{\end{array}}
\def\tep{$B \rar K \ell^+ \ell^-$}
\def\tepm{$B \rar K \mu^+ \mu^-$}
\def\tept{$B \rar K \tau^+ \tau^-$}
\def\ds{\displaystyle}



\def\bos{\lower 0.5cm\hbox{{\vrule width 0pt height 1.2cm}}}
\def\boss{\lower 0.35cm\hbox{{\vrule width 0pt height 1.cm}}}
\def\aaa{\lower 0.cm\hbox{{\vrule width 0pt height .7cm}}}
\def\dol{\lower 0.4cm\hbox{{\vrule width 0pt height .5cm}}}


\title{ {\Large {\bf Pion--baryon coupling constants in light
                     cone QCD sum rules} } }

\author{\vspace{1cm}\\
{\small T. M. Aliev \thanks
{e-mail: taliev@metu.edu.tr}\,\,,
M. Savc{\i} \thanks
{e-mail: savci@metu.edu.tr}} \\
{\small Physics Department, Middle East Technical University} \\
{\small 06531 Ankara, Turkey} }
\date{}

\begin{titlepage}
\maketitle
\thispagestyle{empty}

\begin{abstract}
\baselineskip  0.7cm
We calculate the $\pi\Sigma\Lambda$ and $\pi\Sigma\Sigma$ coupling constants
in light cone QCD sum rules for the structure 
$\sigma_{\alpha\beta} \gamma_5 p^\alpha q^\beta$. A comparison of our results
on these coupling constants with prediction of the SU(3) symmetry is 
presented.
\end{abstract}

\vspace{1cm}
\end{titlepage}

\section{Introduction}
Many strong interaction--processes involve meson--baryon coupling constants
as the main ingredient. The determination of these fundamental quantities
requires information about the physics at large distance. In other words,
for a reliable determination of these parameters we need some
nonperturbative approach. Among all nonperturbative approaches, QCD sum
rules \cite{R1} is one of the most powerful method in studying 
the properties of hadrons. This
method is based on the short distance OPE of vacuum--vacuum correlation
function in terms of condensates.  For the processes involving light mesons
$\pi,~K$ or $\rho$, there is an alternative method to the traditional QCD
sum rules, namely, light cone QCD sum rules \cite{R2}. In this approach the
expansion of the vacuum--meson correlator is performed near the light cone
in terms of the meson wave functions. The meson wave functions are defined
by the matrix elements of non--local composite operators sandwiched between
the meson and vacuum states and classified by their twists, rather
than dimensions of the operators, as is the case in the traditional sum
rules. Many applications of light--cone QCD sum rules can be found in
\cite{R3}--\cite{R11} and references therein. 
 
In this work we use light cone QCD sum rules approach for determination 
of the coupling constants of the pion to the lowest states of the baryon
octet $\Sigma$ and $\Lambda$, $g_{\pi\Sigma\Lambda}$ and 
$g_{\pi\Sigma\Sigma}$. Note that these coupling constants were investigated 
in framework of the QCD sum rules based on pion--to--vacuum matrix element 
in the leading order of the pion momentum $q$ for the structure 
$\not\!q \gamma_5$ in \cite{R12}, where $q$ is the pion momentum. 
The results of this work are currently under debate in literature (see
discussions in \cite{R13} and \cite{R14}).
Moreover in \cite{R15} and \cite{R16} 
it was pointed out that there is
coupling scheme dependence for the structures $\gamma_5$, $\not\!q \gamma_5$,
i.e., dependence on the pseudoscalar or pseudovector forms of the effective
interaction Lagrangian of pion with hadrons 
have been used, while the structure $\sigma_{\mu\nu} \gamma_5$ is shown to
be independent of any coupling schemes. For this reason, in present work we
choose the structure $\sigma_{\mu\nu} \gamma_5 p^\mu q^\nu$, where $p$ and
$q$ are the $\Lambda~(\Sigma)$ and the pion momenta, respectively. It should
be noted that the sum rules for the $\sigma_{\mu\nu} \gamma_5 p^\mu q^\nu$  
structure was derived in \cite{R17} in investigation of $g_{\rho\omega\pi}$
coupling constant.

The paper is organized as follows. In Section 2 we derive sum rules for the
pion--baryon coupling constants $g_{\pi\Sigma\Lambda}$ and
$g_{\pi\Sigma\Sigma}$ for the structure
$\sigma_{\mu\nu} \gamma_5 p^\mu q^\nu$. Section 3 is devoted to the
numerical analysis of the sum rules for $g_{\pi\Sigma\Lambda}$ and
$g_{\pi\Sigma\Sigma}$ and discussion.

\section{Formulation of the pion--baryon sum rule for the 
$g_{\pi\Sigma\Lambda}$ and $g_{\pi\Sigma\Sigma}$}

According to the main philosophy of the QCD sum rules, a 
quantitative estimation for $g_{\pi\Sigma\Lambda}$ and 
$g_{\pi\Sigma\Sigma}$ couplings can be obtained by
matching the representations of a suitable correlator in terms of hadronic 
(physical part) and quark--gluon language (theoretical part). For this
purpose we consider the following two--point correlator function with pion
\bea
\Pi(p,q) = \int d^4x \, e^{ipx} \lla \pi(q) \vel \mbox{\rm T} 
\left[\eta_Y (x) \bar \eta_{\Sigma^+} (0) \right] \ver 0 \rra~,
\eea
where $p$ and $\eta_Y$ are the four--momentum of the hyperon (in our case
$\Lambda^0$ or $\Sigma^0$) and its interpolation current, respectively, 
$\eta_{\Sigma^+}$ is the interpolating current of $\Sigma^+$ and
$q$ is the pion four--momentum.
The interpolating currents for $\Lambda^0$, $\Sigma^0$ and $\Sigma^+$ 
are \cite{R18}
\bea
\eta_{Y^0} &=& \alpha \, \epsilon_{abc} \left[  
\ga u_a^T {\cal C} \gamma_\mu s_b \dr \gamma_5 \gamma^\mu d_c \mp
\ga d_a^T {\cal C} \gamma_\mu s_b \dr \gamma_5 \gamma^\mu u_c \right]~,\nnb \\
\eta_{\Sigma^+}&=& \epsilon_{abc}
\ga u_a^T {\cal C} \gamma_\mu u_b \dr \gamma_5 \gamma^\mu s_c~,
\eea
where $s,~u$ and $d$ are strange, up and down quark fields,
the upper(lower) sign corresponds to $\Lambda^0(\Sigma^0)$ and 
$\alpha=\sqrt{2/3}$ for  $\Lambda^0$ and $\sqrt{2}$ for $\Sigma^0$,
respectively, $a,~b,~c$ are the color indices, ${\cal C}$ is the charge conjugation
operator.   
Saturating correlator (1) with the $Y~(=\Lambda^0$ or $\Sigma^0$) 
and $\Sigma^+$ states in the phenomenological part, we get 
\bea     
\Pi=       
\frac{\lla \pi(q) Y \ve \Sigma^+ \rra \, 
\lla \Sigma^+(p+q) \ve \bar \eta_{\Sigma^+} \ve 0 \rra
\, \lla 0\ve \bar \eta_Y \ve Y(p) \rra}
{(p^2-m_Y^2) \left[(p+q)^2 - m_{\Sigma^+}^2\right]} +
\mbox{\rm high. reson.}
\eea     
The matrix elements in Eq. (3) are defined in the following way
\bea
\lla 0 \vel \eta_Y(x) \ver Y(p) \rra &=& \lambda_Y u(p)~, \nnb \\
\lla \Sigma^+(p+q) \vel \bar \eta_{\Sigma^+} \ver 0 \rra &=& 
\lambda_\Sigma \bar u(p+q)~,\nnb \\
\lla \pi(q) Y(p)\ve \Sigma^+(p+q) \rra &=& -
g_{Y\Sigma^+\pi^-} \, \bar u(p) \gamma_5 u(p+q)~.
\eea
Substituting Eq. (4) in Eq. (3), and choosing the structure
$i\, \sigma_{\alpha\beta} p^\alpha q^\beta \gamma_5$ 
for the physical part of Eq. (1), we get
\bea
\Pi^{phys}= - \frac{g_{\pi^-Y\Sigma^+} \lambda_Y \lambda_{\Sigma^+}}
{(p^2-m_Y^2) \left[(p+q)^2 - m_\Sigma^2\right]}
+ \mbox{\rm higher resonances}~.
\eea
Let us now consider the theoretical part of the correlator (1). From
this correlator we have (we present only the terms which give contributions
to the above--mentioned Lorentz structure)
\bea
\Pi &=& - \alpha \int d^4x\,e^{ipx} \Big\{ - \gamma_5 \gamma_\mu 
\gamma_5 \gamma_\varphi \gamma_\rho\,
{\cal C}{\cal S}^T {\cal C}^{-1} \gamma_\mu {\cal S}^s \gamma_\rho
\gamma_5 \lla \pi \vel \bar u \gamma_5 \gamma_\varphi d \ver 0 \rra \nnb \\
&&+\frac{1}{2} \gamma_5 \gamma_\mu \sigma_{\alpha\beta} \gamma_\rho\,
{\cal C}{\cal S}^T {\cal C}^{-1} \gamma_\mu {\cal S}^s \gamma_\rho
\gamma_5 \lla \pi \vel \bar u \sigma_{\alpha\beta} d \ver 0 \rra \nnb \\
&& \pm \Big[ - \gamma_5 \gamma_\mu {\cal S} \gamma_\rho \,{\cal C} 
(\gamma_5 \gamma_\varphi)^T 
{\cal C}^{-1} \gamma_\mu {\cal S}^s \gamma_\rho \gamma_5 
\lla \pi \vel \bar u \gamma_5 \gamma_\varphi d \ver 0 \rra \nnb \\
&&+ \frac{1}{2} \gamma_5 \gamma_\mu {\cal S} \gamma_\rho \,{\cal C}
\sigma_{\alpha\beta}^T {\cal C}^{-1} \gamma_\mu {\cal S}^c \gamma_\rho \gamma_5
\lla \pi \vel \bar u \sigma_{\alpha\beta}d \ver 0\rra\Big] \Big\} ~,
\eea
where upper(lower) sign corresponds to $\Lambda(\Sigma)$ case and
$\alpha=\sqrt{2/3}$ for $\Lambda$ while $\alpha=\sqrt{2}$ for $\Sigma$.
Here ${\cal S}$ and ${\cal S}^s$ are the propagators containing both
perturbative and nonperturbative contributions, respectively. Here we
present the explicit form of $i {\cal S}^s(x)$  
\bea
\lefteqn{
i {\cal S}^s(x) = \frac{i}{2 \pi^2} \frac{\not\!x}{x^4} -
\frac{m_s}{4\pi^2} \frac{1}{x^2} - \frac{1}{12} 
\la \bar s s \ra \ga 1 - \frac{i m_s}{4} \not\!x \dr} \nnb \\
&& - \frac{1}{192} m_0^2 \la \bar s s \ra 
 \ga 1 - \frac{i m_s}{6} \not\!x \dr -
i g_s \frac{1}{16 \pi^2} \int_0^1 du \left\{
\frac{\not\!x}{x^2} \sigma_{\alpha\beta} G^{\alpha\beta}(ux) -
4i \frac{x_\alpha}{x^2}G^{\alpha\beta} \gamma_\beta \right\}~,
\eea
where $m_s$ is the mass of the strange quark and $G_{\alpha\beta}$
is the gluon field strength tensor.
The form of ${\cal S}$ can be obtained from Eq. (7) by making the
replacements $\la \bar s s \ra  \rar \la \bar q q \ra$ and 
$m_s \rar 0$. From Eq. (6) we observe that, in calculation of the
correlator function in QCD, the matrix element of the nonlocal operators
between the vacuum and pion states are needed. These matrix elements define
the two particle pion wave functions and up to twist four they can be 
written as \cite{R6,R7}
\bea
\lla \pi(q) \vel \bar d  \gamma_\mu \gamma_5 u  \ver 0 \rra &=&
- i f_\pi q_\mu \int_0^1 du \, e^{iuqx} \big[ \varphi_\pi(u) + x^2 g_1(u) \big]
\nnb \\
&&+f_\pi \ga x_\mu - \frac{x^2 q_\mu}{qx} \dr
\int_0^1 du\, e^{iuqx} g_2(u)~, \nnb \\ \nnb \\
\lla \pi(q) \vel \bar d (x) \sigma_{\alpha\beta}\gamma_5 u(0) \ver 0 \rra &=&
i \ga q_\alpha x_\beta - q_\beta x_\alpha \dr 
\frac{f_\pi m_\pi^2}{6(m_u+m_d)}
 \int_0^1 du \,e^{iuqx} \varphi_\sigma(u)~.
\eea
Using Eqs. (6), (7) and (8) we get the following result 
for theoretical part (for the
structure $i \sigma_{\alpha\beta} x_\alpha q^\beta \gamma_5$)
\bea
\lefteqn{ 
\Pi^{theor} =} \nnb \\ 
&& - \alpha f_\pi \int_0^1 dx \,e^{ipx} \Bigg\{ 
\Bigg[ (1 \pm 1) \frac{m_s}{2 \pi^4 x^6} + 
(-\lambda \pm \sigma) \ga \frac{1}{6 \pi^2 x^4} +
\frac{m_0^2}{96 \pi^2 x^2} \dr \Bigg] 
\int_0^1 du\, e^{iuqx} \varphi_\pi(u) \nnb \\
&&+\Bigg[ (1 \mp 1) \frac{\mu_\pi}{6 \pi^4 x^6}
+ (1 \pm 1) \frac{\mu_\pi \lla g^2 G^2 \rra}{2304 \pi^4 x^2}+
(1 \mp 1) \frac{m_s \mu_\pi \lla \bar s s \rra}{144 \pi^2 x^2}
\pm \frac{m_s \mu_\pi \sigma}{72 \pi^2 x^2} 
\Bigg] \int_0^1 du\, e^{iuqx} \varphi_\sigma(u) \nnb \\
&& +\Bigg[ \frac{1}{6 \pi^2 x^2} (-\lambda \pm \sigma) 
+ (1 \pm 1) \frac{m_s}{2 \pi^4 x^4} \Bigg]
\int_0^1 du\, e^{iuqx} \left[ g_1(u) + G_2 (u) \right] \Bigg\}~,
\eea
where 
\bea
G_2(u) &=& - \int_0^u g_2(v) dv~, \nnb \\
\lambda &=& \la \bar q q \ra - \la \bar s s \ra~, \nnb \\
\sigma &=& \la \bar q q \ra + \la \bar s s \ra~. \nnb
\eea
Our next task is to perform integration over $x$ and perform double Borel
transformation in Eq. (9) with respect to the variables $p^2$ and $(p+q)^2$,
in order to get an answer for the theoretical part of the sum rules. As an
example, let us demonstrate on one of the terms in Eq. (9) how integration
over $x$ and double Borel transformation can be carried. Consider the
following term
\bea
\int du \varphi_\pi(u) \int d^4 x \frac{e^{i(p+qu)x}}{x^2} i
\sigma_{\alpha\beta} x_\alpha q_\beta~. \nnb
\eea
In performing the $x$ integration, we will make use of the
formula \cite{R19}
\bea
\int \frac{d^4 x}{(x^2)^n} e^{ipx} = 
\frac{i (-1)^n 2^{4-2n} \pi^2}{\Gamma(n-1) \Gamma(n)}
(p^2)^{n-2} ln(-p^2) + {\cal P}_{n-2}~,~~~(n \ge 2 )~, \nnb    
\eea
where ${\cal P}_{n-2}$ is a polynomial of power $n-2$. However this polynomial
is inessential, since it vanishes after double Borel transformation. Hence,
disregarding this polynomial we have
\bea
\int du \varphi_\pi(u) \int d^4 x \frac{e^{i(p+qu)x}}{x^2} 
i \sigma_{\alpha\beta} x_\alpha q_\beta &=&
\int du \varphi_\pi(u) \ga i \frac{\partial}{\partial p_\alpha}
\frac{\partial}{\partial p_\rho} \frac{\partial}{\partial p_\rho} \dr
\int d^4 x \frac{e^{i(p+qu)x}}{x^4} i \sigma_{\alpha\beta} q_\beta \nnb \\
&=& 8 \int du \varphi_\pi(u)\frac{P_\alpha}{P^4} i \sigma_{\alpha\beta} q_\beta ~, \nnb
\eea
where $P=p + qu$. The last step in this calculation is performing double
Borel transformation over the variables $p^2$ and $(p+q)^2$ to the
expression
\bea
\int du \varphi_\pi(u) \frac{1}{\left[ \ga p+qu \dr^2 \right]^2}~. \nnb
\eea
Rewriting $(p+qu)^2 = p^2 \bar u + u (p+q)^2 $ (here the pion mass is
neglected) and using the exponential representation for the denominator,
\bea
\frac{1}{\left[ p^2 \bar u + u (p+q)^2 \right]^2} = 
\int_0^\infty d \alpha \alpha e^{-\alpha \left[ p^2 \bar u + 
u (p+q)^2 \right]}~, \nnb
\eea
we have 
\bea
\int du \varphi_\pi(u) \frac{1}{\left[ \ga p+qu \dr^2 \right]^2} =
\int du \varphi_\pi(u) \int d \alpha \alpha e^{-\alpha \left[ p^2 \bar u + 
u (p+q)^2 \right]}~. \nnb
\eea
The double Borel transformation over the variable $p^2$ and $(p+q)^2$ is
done with the help of the following general formula
\bea
{\cal B}_{p^2}^{M_1^2} {\cal B}_{(p+q)^2}^{M_2^2}
\frac{\Gamma(n)}{\left[ - \bar u p^2 - (p+q)^2 u \right]^n} =
(M^2)^{2-n} \delta (u-u_0)~,\nnb
\eea
where
\bea
M^2 = \frac{M_1^2 M_2^2}{M_1^2+M_2^2}~~~~~\mbox{\rm and} ~~~~~ 
u_0 = \frac{M_1^2}{M_1^2+M_2^2}~,\nnb
\eea
and $M_1^2$ and $M_2^2$ are the Borel parameters. 
Using this expression and performing the integrations over the variables 
$\alpha$ and $u$, we finally get
\bea
\int du \varphi_\pi(u) \int d^4 x \frac{e^{i(p+qu)x}}{x^2} i
\sigma_{\alpha\beta} x_\alpha q_\beta = 8 i \sigma_{\alpha\beta} 
p_\alpha q_\beta \varphi_\pi(u_0)~. \nnb
\eea
All other terms can be calculated similarly and for
the theoretical part it follows from Eq. (9) that
\bea
\lefteqn{
\Pi^{theor} =} \nnb \\ 
&&- \alpha f_\pi \Bigg\{ \varphi_\pi(u_0) \Bigg[ 
(1 \pm 1) \frac{m_s}{8 \pi^2} M^4 f_1(s_0/M^2)
- \frac{1}{3} ( -\lambda \pm \sigma ) M^2 f_0(s_0/M^2)
+\frac{m_0^2}{12}( -\lambda\pm\sigma ) \Bigg] \nnb \\
&& + \mu_\pi \varphi_\sigma(u_0) \Bigg[ (1 \mp 1) \frac{1}{24 \pi^2}
M^4 f_1(s_0/M^2) + (1 \pm 1) \frac{8}{2304} 
\frac{\lla g^2 G^2 \rra}{\pi^2}
+ (1 \mp 1) \frac{1}{18} m_s \la \bar s s \ra
\pm \frac{1}{9} m_s \sigma \Bigg] \nnb \\
&&+ \left[ g_1(u_0) + G_2(u_0) \right] \Bigg[ 
-(1 \pm 1) \frac{1}{\pi^2} m_s M^2 f_0(s_0/M^2) + 
\frac{4}{3} \ga -\lambda \pm \sigma \dr \Bigg] \Bigg\}~,
\eea
where the function 
\bea
f_n(s_0/M^2)=1-e^{-s_0/M^2}\sum_{k=0}^n \frac{(s_0/M^2)^k}{k!}~, \nnb
\eea
is the factor used to
subtract the continuum, which is modeled by the dispersion integral in the
region $s_1,~s_2 \ge s_0$, $s_0$ being the continuum threshold (obviously
the continuum thresholds for the $\Lambda$ and $\Sigma$ channels are 
different). Since masses of $\Lambda$
and $\Sigma$ are very close to each other, we can choose
$M_1^2$ and $M_2^2$ to be equal to each other, i.e., $M_1^2 = M_2^2 =2 M^2$,
from which it follows that $u_0=1/2$. 

Performing double Borel transformation over the variables $p^2$ and
$(p+q)^2$ in the physical part (5) and then equating the the obtained result
to Eq. (10), we get the sum rules for $g_{\pi\Lambda\Sigma}$ and
$g_{\pi\Sigma\Sigma}$ coupling constants  
\bea
g_{Y\Sigma^+\pi^-} \lambda_Y \lambda_{\Sigma^+} = 
e^{m^2/M^2} \Pi^{theor}~,
\eea
where $m\approx m_\Lambda \approx m_\Sigma$.

From Eq. (11) it follows that in determining the strong coupling constants 
$g_{\pi\Lambda\Sigma}$ and $g_{\pi\Sigma\Sigma}$ the experimentally
undetermined residues $\lambda_\Lambda$ and $\lambda_\Sigma$ need to be
eliminated from sum rules. The residues $\lambda_\Lambda$ and
$\lambda_\Sigma$ are determined from corresponding mass sum rules for the 
$\lambda$ and $\Sigma$ hyperons \cite{R18,R20} as follows
\bea
\vel \lambda_\Lambda \ver^2 e^{-m_\Lambda^2/M^2} 32 \pi^4 &=& M^6
f_2(s_0^\Lambda/M^2) + \frac{2}{3} a m_s (1 - 3 \gamma) M^2
f_0(s_0^\Lambda/M^2)\nnb \\
&&+ b M^2 f_0(s_0^\Lambda/M^2) + \frac{4}{9}
a^2 (3+4 \gamma)~,\\
\vel \lambda_\Sigma \ver^2 e^{-m_\Sigma^2/M^2} 32 \pi^4 &=& M^6
f_2(s_0^\Sigma/M^2) - 2 a m_s (1+\gamma)M^2 f_0(s_0^\Sigma/M^2) \nnb \\
&&+b M^2 f_0(s_0^\Sigma/M^2) + \frac{4}{3} a^2~,
\eea
where
\bea
a &=& - 2 \pi^2 \la \bar q q \ra ~, \nnb \\
b &=& \frac{\alpha_s \la G^2 \ra }{\pi} \simeq 0.012~GeV^4~, \nnb \\
\gamma &=& \frac{\la \bar s s \ra}{\la \bar q q \ra} - 1 
\simeq  - 0.2~, \nnb
\eea
and the functions $f_0(x),~f_1(x)$ are presented just after Eq. (10).
The ratio of the Eqs. (11), (12) and (13) gives
\bea
g_{\pi\Lambda\Sigma} &=& e^{m^2/M^2} \frac{\Pi^{theor}}
{\lambda_\Lambda \lambda_\Sigma}~, \\
g_{\pi\Sigma\Sigma} &=& e^{m^2/M^2} \frac{\Pi^{theor}}
{\lambda_\Sigma^2}~.
\eea
The main reason why we consider the above--mentioned ratio
rather than the individual sum rules themselves (i.e., first determine 
$\lambda_\Lambda$ and $\lambda_\Sigma$ independently from Eqs. (12) and (13)
and substitute their obtained values in Eq. (11)) is that the sum
rules obtained from these ratios are more stable as is similar to the baryon
mass sum rules case. In addition to that the 
uncertainties coming from various parameters such as quark condensate, 
$m_0^2$, continuum threshold $s_0$ and Borel parameter, are reduced.

\section{Numerical analysis}
Now we are ready to perform the numerical analysis. The main nonperturbative
input parameters in the sum rules (11) are the pion wave
functions. In our calculations we have used the set of wave functions
proposed in \cite{R6}. The explicit expressions of the wave functions are
\bea
\varphi_\pi(u,\mu) &=& 6 u \bar u \left[ 1 + a_2(\mu) C_2^{3/2} (2 u -1 ) +
a_4(\mu) C_4^{3/2} (2 u -1 )\right]~, \nnb \\
\varphi_\sigma(u,\mu) &=& 6 u \bar u \left[ 1 + C_2 \frac{3}{2}
\left[5(u-\bar u)^2 -1 \right] + C_4 \frac{15}{8} 
\left[21(u-\bar u)^4 - 14 (u-\bar u)^2 + 1\right]\right]~, \nnb \\ 
g_1(u,\mu) &=& \frac{5}{2} \delta^2(\mu)\bar u^2 u^2 +
\frac{1}{2} \varepsilon(\mu) \delta^2(\mu) \Bigg[ u \bar u
(2 + 13 u \bar u) \nnb \\
&& + 10 u^3 \ln u \ga 2 - 3u + \frac{6}{5} u^2 \dr +
10 \bar u^3 \ln \bar u \ga 2 - 3\bar u +
\frac{6}{5}\bar u^2 \dr\Bigg]~, \nnb \\
G_2(u,\mu) &=& \frac{5}{3} \delta^2(\mu) \bar u^2 u^2~,
\eea
where $\bar u = 1-u$, $C_2^{3/2}$ and $C_4^{3/2}$ are the Gegenbauer 
polynomials defined as
\bea
C_2^{3/2} (2 u -1 ) &=& \frac{3}{2} \left[ 5(2 u-1)^2 +1 \right]~,\nnb \\
C_4^{3/2} (2 u -1 ) &=& \frac{15}{8} \left[ 21 (2 u -1 )^4 -
14 (2 u -1 )^2 +1 \right] ~,
\eea
and $a_2(\mu=0.5~GeV)=2/3$, $a_4(\mu=0.5~GeV)=0.43$. The parameters
$\delta^2(\mu=1~GeV)=0.2~GeV^2$ \cite{R21} and $\epsilon(\mu=1~GeV)=0.5$
\cite{R6}. Furthermore $f_\pi=0.132~GeV$, $\mu_\pi(\mu=1~GeV)=1.65$,
$\la \bar s s \ra = 0.8 \la \bar q q \ra$ and
$\la \bar q q \ra\ve_{\mu=1~GeV}=-(0.243)^3~GeV^3,~s_0 = s_0^\Lambda \simeq
s_0^\Sigma = (3.0 \pm 0.2)~GeV^2$. Moreover all further calculations
are performed at $u=u_0=1/2$.

Having fixed the input parameters, one
must find the range of values of $M^2$ over which the sum rule
is reliable. The lowest possible value of $M^2$ is determined by
the requirement that the terms proportional to the highest inverse power of
the Borel parameters stay reasonable small. The upper bound of $M^2$ is
determined by demanding that the continuum contribution is not too large.
The interval of $M^2$ which satisfies both conditions is
$1 ~GeV^2 < M^2 < 2.5~GeV^2$. The analysis of the sum rules (14) and (15) shows that
the best stability is achieved in the region of $M^2$, 
$1.4~GeV^2 < M^2 < 1.8~GeV^2$. This leads to the following result for the
coupling constants 
\bea
g_{\pi\Lambda\Sigma} &=& 5.3 \pm 1.8 \nnb \\
g_{\pi\Sigma\Sigma}  &=& 12.5 \pm 4.5~,
\eea
in which the errors coming from the quark condensate (which varies in the
region $-(0.24~GeV)^3$ and $-(0.26~GeV)^3$), $m_0^2$ parameter
(which varies in the region $0.6~GeV^2$ and $1.4~GeV^2$), variation of the
Borel parameter and change of the continuum threshold $s_0$ are all taken
into consideration. Our calculation shows that the main error comes from
uncertainties of the quark condensate. The central values of the coupling
constants are obtained at $m_0^2=0.8~GeV^2$, 
$\la \bar q q \ra = - (0.243~GeV)^3 $ and $s_0=3.0~GeV^2$.

At this point we would like to stress that the above--mentioned strong
coupling constant have been analyzed using the individual sum rules
themselves by first determining $\lambda_\Lambda$ and $\lambda_\Sigma$
independently from Eqs. (12) and (13) and substituting their obtained values
in Eq. (11). The results predicted by this approach are close to the ones 
presented in Eq. (18), however our calculations show that 
the ratio sum rules prediction is more stable and reliable.   
  
Here it should be noted that since the phase of the coupling constants can not be
predicted by the sum rules, we take into consideration
the magnitudes of these coupling constants to be able to compare them with
the predictions of other approaches.  

Let us now compare our results on the coupling constants 
$g_{\pi\Lambda\Sigma}$ and $g_{\pi\Sigma\Sigma}$ with SU(3) symmetry
prediction. As is known, SU(3) symmetry predicts 
\bea
G_{\pi\Lambda\Sigma} &=& \frac{2}{\sqrt{3}} (1-\alpha)\, G_{NN\pi} ~,\nnb \\
G_{\pi\Sigma\Sigma}  &=& 2 \alpha \,G_{NN\pi} ~,
\eea
where $\alpha = F/(F+D)$ (see for example \cite{R22}). Exact SU(3) symmetric
analysis of pion--baryon coupling gives $F/D \simeq 0.58$
\cite{R22} (exact SU(6) symmetry predicts 
$F/D=2/3$). It follows from Eqs. (19) that
\bea
R = \frac{G_{\pi\Sigma\Sigma}}{G_{\pi\Lambda\Sigma}} = 
\frac{\sqrt{3}\alpha}{1-\alpha} \approx 1~,
\eea
while our analysis yields $R\simeq 2$. It follows from a comparison of these 
results that SU(3) symmetry is broken significantly.
The pion--baryon couplings predicted in \cite{R12} are not
presented here since we have already noted that the results of
this work is currently under debate in literature. 

In conclusion we have calculated the strong coupling constants of pion with 
$\Lambda$ and $\Sigma$ hyperons and found out that our results differ
significantly from that of the SU(3) symmetry prediction.

\end{document}